\documentclass[12pt]{article}
\usepackage{enumitem}
\usepackage{amsfonts}
\usepackage{amsmath}
\usepackage{graphicx}
\usepackage{subfigure}
\usepackage{tabularx}
\usepackage{fancyhdr}
\usepackage{amssymb}
\usepackage{amsmath}
\usepackage{mathrsfs}
\usepackage{float}
\usepackage{verbatim}
\usepackage{algorithm}
\usepackage{algpseudocode}
\usepackage{mathtools}
\usepackage[authoryear]{natbib}
\usepackage{array,graphicx}
\usepackage{enumerate}
\usepackage{graphicx}
\usepackage{multirow}
\usepackage{subfigure}
\usepackage{leftidx}
\usepackage{setspace}
\usepackage{tikz}
\usetikzlibrary{shapes.geometric, arrows}
\usepackage[titletoc,toc,title]{appendix}
\usepackage{amsmath}
\usepackage{graphicx}

\usepackage{colortbl}
\usepackage{booktabs}
\usepackage{adjustbox}
\usepackage{relsize}
\usepackage{booktabs}
\usepackage{diagbox}
\usepackage{extarrows}
\usepackage{multirow}

\newcommand{\R}{\mathbb{R}}

\usepackage{setspace}


\begin{document}
\title{Portfolio selection under two-factor stochastic volatility and transaction costs with option-implied utility and deep learning}
\author{{Dong Yan$^{a}$, Ke Zhou$^{a}$, Zirun Wang$^{a}$, Xin-Jiang He$^{b,c,}$\thanks{Corresponding author: xinjiang@zjut.edu.cn.}}\\
\small\it a. School of Statistics, University of International Business and Economics, Beijing, China.\\
{\small\it b. School of Economics, Zhejiang University of Technology, Hangzhou, China.}\\
{\small\it c. Institute for Industrial System Modernization, Zhejiang University of Technology, Hangzhou, China.}
}
\date{}
\maketitle

\begin{abstract}
In this paper, we investigate a portfolio selection problem with transaction costs under a two-factor stochastic volatility structure, where volatility follows a mean-reverting process with a stochastic mean-reversion level. The model incorporates both proportional exogenous transaction costs and endogenous costs modeled by a stochastic liquidity risk process. Using an option-implied approach, we extract an S-shaped utility function that reflects investor behavior and apply its concave envelope transformation to handle the non-concavity. The resulting problem reduces to solving a five-dimensional nonlinear Hamilton–Jacobi–Bellman equation. We employ a deep learning-based policy iteration scheme to numerically compute the value function and the optimal policy. Numerical experiments are conducted to analyze how both types of transaction costs and stochastic volatility affect optimal investment decisions.
\end{abstract}


\section{Introduction}

Numerous scholars have contributed to the study of portfolio selection, with seminal works such as the single-period mean-variance framework introduced by \cite{markowitz1952portfolio} and the continuous-time expected utility maximization approach developed by \cite{merton1975optimum}. These foundational models have laid the groundwork for subsequent research in the field. However, classical models, along with much of the literature built upon them, often operate under the assumption of an idealized, complete market. In reality, financial markets are plagued by frictions, moral hazard, and information asymmetry. To better reflect empirical conditions, this paper focuses on the more realistic setting of an incomplete market, with particular emphasis on transaction costs. Based on their nature, we classify these costs into two categories: exogenous and endogenous transaction costs.

Exogenous transaction costs refer to explicit, directly observable fees incurred per trade, typically including taxes, commissions, and other charges in various forms. Due to the highly heterogeneous and complex nature of such fees, it is difficult to capture exogenous costs using a single realized expression. As a result, numerous studies have explored different functional forms of these costs within the Markowitz framework. Proposed models range from simple fixed transaction costs \citep{patel1982simple}, linear transaction costs \citep{best2005algorithm}, nonlinear transaction costs \citep{olivares2018robust, chellathurai2005dynamic}, to conditionally linear transaction costs that incorporate minimum charge thresholds \citep{baule2010optimal, lobo2007portfolio}. Among these, the proportional transaction cost model remains the most widely adopted and extensively studied specification in the literature. Proportional transaction costs can be further divided into two types. Although both assume a fixed cost rate, some studies model costs as proportional to the total trade value \citep{davis1990portfolio}, while a larger body of work focuses on costs proportional to the change in asset value resulting from the transaction \citep{li2018transaction, peng2008mean, dybvig2020mean}. The latter approach has gained broader acceptance and is generally considered more reflective of real-world market conditions.

While the aforementioned studies are primarily set within the discrete-time Markowitz framework, this approach is inherently static and single-period. It thus fails to accommodate dynamic decision-making that adjusts to evolving market environments. Moreover, a naive extension to a multi-period setting introduces time inconsistency, violating Bellman’s principle of optimality. To address these limitations, this paper adopts Merton’s continuous-time framework and formulates the problem as one of expected utility maximization over time. Several studies on exogenous transaction costs have been developed within the Merton framework \citep{choi2007algorithm, dai2008penalty, dai2010continuous}, most of which also employ proportional transaction costs. However, research in continuous time is considerably more challenging than in the single-period case, as it generally requires solving a Hamilton-Jacobi-Bellman (HJB) equation, a process that entails significant mathematical and computational difficulties.

Distinct from exogenous transaction costs, endogenous transaction costs originate from within the transaction process itself and are directly shaped by the behaviors of participating agents. A prominent example is the liquidity cost, which stems from liquidity risk, a pervasive feature of real financial markets. Liquidity risk significantly influences asset pricing, making its integration essential both in derivative valuation and in optimal portfolio selection strategies. As a result, effectively modeling liquidity risk has become an important research focus, spurring investigations into its financial properties and empirical characteristics.

The literature has approached liquidity risk modeling from multiple perspectives. Some studies characterize it through the bid-ask spread \citep{bangia2008modeling, abensur2022improving}, while others extend traditional Value-at-Risk (VaR) models by incorporating regularization penalty terms \citep{al2021multivariate, botha2008portfolio}. Another strand of research constructs market-impact functions to capture liquidity effects \citep{ly2007model, caccioli2016liquidity, ha2020algorithmic}. Additionally, a number of studies model liquidity risk directly as a stochastic process \citep{feng2014option, feng2016importance, tian2013efficient, abensur2019stochastic, pasricha2023exchange}.

Notably, \cite{feng2014option} formalized this approach by proposing that the liquidity discount factor follows a stochastic process, while also modeling market liquidity as a mean-reverting stochastic process. These two interrelated processes jointly characterize liquidity risk. In subsequent work, \cite{feng2016importance} empirically demonstrated the superiority of this stochastic liquidity modeling approach in the context of European option pricing. More recently, \cite{pasricha2023exchange} extended the framework by introducing a more generalized correlation structure among the driving Brownian motions, further generalizing the model originally proposed by \cite{feng2014option}.

Beyond transaction costs, the accurate modeling of volatility is equally critical. Early studies, notably the Black-Scholes option pricing model \citep{black1973pricing}, were built on the assumption of constant volatility. However, this assumption has been widely shown to fail in capturing the dynamic behavior of real-world market volatility. In response, substantial research efforts have been directed toward developing more realistic volatility models. Among these, the stochastic volatility model proposed by \cite{heston1993closed} has gained considerable prominence. In this framework, the volatility of the underlying asset is modeled as a stochastic process governed by Cox-Ingersoll-Ross (CIR) dynamics. This formulation not only enables the derivation of semi-analytical solutions for European option prices but also guarantees the non-negativity and mean-reverting property of volatility. Thanks to these desirable features, the Heston model has been widely adopted in numerous portfolio selection studies within the Merton framework \citep{fouque2017portfolio, kraft2005optimal, zeng2013stochastic}. Furthermore, building on the Heston model, \cite{he2021closed} introduced a refinement to overcome its limitation in adequately capturing nonlinear mean reversion. Their approach assumes that the mean reversion level itself follows a stochastic process, thereby extending the original Heston model into a two-factor stochastic volatility framework. This enhanced model is adopted in the present study.

Furthermore, the expected utility maximization problem in the Merton framework requires the explicit specification of a utility function. A significant body of literature employs market data from stocks or options to estimate investor utility functions. Researchers have primarily pursued two approaches: some have developed nonparametric methods to recover state-dependent risk aversion functions from observed market data \citep{ait2000nonparametric, jackwerth2000recovering, rosenberg2002empirical}, while others have calibrated preference parameters by testing parametric asset pricing models against various hypothesized utility specifications \citep{garcia2003empirical, bliss2004option}. Among these, the option-implied methodology has gained considerable prominence for extracting risk aversion functions. This approach infers the risk-neutral probability density function (PDF) from the implied volatility smile of options. The estimated utility function is then applied to transform this risk-neutral PDF into a subjective PDF, after which the optimal utility function is selected using statistical techniques. A major advantage of this method is that it facilitates clear comparative analysis across different utility functions by maintaining a consistent structural form during estimation. To date, applications have been largely confined to classical utility functions, as exemplified by \cite{bliss2004option}, although \cite{Kang2006} extended the approach to include HARA-type and certain composite utility functions. This paper substantially broadens the scope by incorporating the S-shaped utility function from prospect theory \citep{kai1979prospect}, which is widely acknowledged for its descriptive accuracy of real investor behavior. Notably, the application of the option-implied approach to recover such an S-shaped utility specification remains scarce \citep{driessen2007empirical}, positioning our study as a meaningful contribution toward addressing this gap in the literature.

This paper investigates the continuous-time portfolio selection problem within the Merton framework, formulated as an expected utility maximization problem. We adopt an S-shaped utility function and account for the effects of both exogenous and endogenous transaction costs. To model volatility, we develop a two-factor stochastic model that captures its inherent randomness. This leads to the formulation of an HJB equation, which must be solved numerically. The core computational challenge thus reduces to solving a five-dimensional nonlinear HJB equation. Given the high dimensionality and strong nonlinearity of this equation, we employ Physics-Informed Neural Networks (PINNs) \citep{raissi2019physics} combined with a deep learning-based policy iteration scheme to obtain accurate and reliable numerical solutions \citep{alla2015efficient, forsyth2007numerical, kerimkulov2020exponential}. Additionally, the non-concave regions of the S-shaped utility function pose significant optimization difficulties. To overcome this, we primarily utilize the concave envelope transformation technique, which is a well-established approach in the literature for handling non-concave utilities \citep{liang2020classification, davey2024deep, dai2022nonconcave, reichlin2013utility, bian2019utility}. This treatment significantly improves the stability and tractability of our numerical optimization procedure.

The remainder of this paper is structured as follows. Section 2 develops the wealth dynamics model, incorporating both exogenous and endogenous transaction costs as well as a two-factor stochastic volatility specification. This formulation leads to the expected utility maximization objective and the corresponding five-dimensional HJB equation. In Section 3, we apply an option-implied methodology to comparatively analyze and calibrate the parameters of the S-shaped utility function. Section 4 implements the concave envelope transformation for the utility function and proceeds with the numerical solution of the HJB equation. This section further conducts a comprehensive numerical analysis, providing illustrative examples to investigate the sensitivity of optimal investment strategies to key parameters, particularly the effects of varying exogenous and endogenous transaction costs and stochastic volatility.

\section{The mode formulation}
In this section, we formulate a dynamic portfolio selection model based on a two-factor stochastic volatility model \citep{Lin2024}, where stock trading incurs both exogenous transaction costs (proportional transaction costs) and endogenous transaction costs (implicitly resulting from liquidity risks). Consider a financial market consisting of two types of assets: risky stocks and risk-free money accounts with an interest rate $r$. We assume that the market is illiquid and that the stochastic processes of stock price, volatility and liquidity risk are established under a physical measure.

To model how liquidity influences equity prices, a market liquidity variable $L$ is defined, which follows a mean-reverting Ornstein-Uhlenbeck stochastic process as detailed in \cite{feng2014option}:
\begin{equation}\label{dL}
dL=\alpha(\theta_L-L)dt+\sigma_L dB^L,
\end{equation}
where $\alpha$ is the mean-reversion speed, $\theta_L$ is the mean-reversion level, and $\sigma_L$ is the volatility of market liquidity.

Moreover, the incorporation of proportional transaction costs must address their intrinsic relationship with liquidity risk. From a theoretical perspective, the rise in exogenous transaction costs diminishes trading incentives by compressing profit margins, thereby increasing market illiquidity. We therefore modify the liquidity risk model in Eq. (\ref{dL}) by introducing a mean-reversion level that increases with the transaction cost rate:
\begin{equation}
\theta_L(L)=\hat{\theta}_L+\lambda_{TC}\cdot \kappa_{TC} \cdot L^\xi,
\end{equation}
where $\hat{\theta}_L$ denotes the current illiquidity level unaffected by transaction costs, $\lambda_{TC}$ represents the sensitivity coefficient quantifying the marginal impact of transaction costs on illiquidity, $\kappa_{TC}$ is the proportional transaction costs rate, and $\xi\in(0, 1)$ governs the curvature of the power function. Such a concave function is adopted for its flexibility in modeling how transaction costs nonlinearly increase liquidity risk through curvature adjustments.

Then the price of the underlying asset $S$, stochastic volatility $v$ with its stochastic mean-reversion level $\theta$, and liquidity risk $L$ satisfy the following stochastic differential equations: 
\begin{equation}\label{dynamics-stock}
\begin{cases}
        d S_{t}= \mu S_t d t+\sqrt{v}S_t d B^{S}_t+\beta L_tS_t d B^\gamma_t,\\
        dv_t=\kappa(\theta_t-v_t)dt+\sigma_1\sqrt{v_t}d B^v_t,\\
        d\theta_t=\lambda(\eta-\theta_t)d t+\sigma_2\sqrt{\theta_t} d B^{\theta}_t,\\
         d L_{t}=\alpha(\hat{\theta}_L+\lambda_{TC}\cdot \kappa_{TC} \cdot L^\xi_t-L_t) d t+\sigma_L d B^L_t,
\end{cases}
\end{equation}
where $\mu$ is the drift,  and the strictly positive parameter $\beta$ measures the sensitivity to the level of market liquidity of the asset price. The stochastic volatility $v$ and its mean-reversion level $\theta$ follow mean-reversion processes with their respective mean-reversion levels and speeds. $\sigma_1$ and $\sigma_2$ correspond to the volatility of volatility and the volatility of the stochastic mean-reversion level, respectively. $B^S$, $B^\gamma$, $B^v$, $B^\theta$ and $B^L$ are correlated Wiener processes with coefficients specified as: $d B^S_t d B^v_t=\rho_1 dt$, $d B^S_t d B^\theta_t=\rho_2 dt$, $d B^v_t d B^\theta_t=\rho_3 dt$, $d B^S_t d B^\gamma_t=\rho_4dt$, $ d B_t^S d B^L_t=\rho_5 d t$, and $d B^\gamma_t d B^L_t=\rho_6 d t$.

Now consider an investor who invests an initial endowment of $W_0$, allocating a fraction $\omega(t)\in [0, 1]$ to stocks and the remainder to a bank account earning the risk-free rate $r$. To avoid excessive costs from continuous trading, we instead hedge the portfolio in a non-infinitesimal time step of length $\delta t$. The associated exogenous transaction costs are assumed to be proportional to the monetary value of the traded stocks. Consequently, the net change in the investor’s wealth over one time step is given by:
\begin{equation}
\delta W_t=[rW_t+(\mu-r)\omega_t W_t] \delta t+\beta\omega_t L_t W_t \delta B_t^\gamma+\omega \sqrt{v_t}W_t\delta B_t^S-\kappa_{TC}S_t|\nu_t|,
\end{equation}
where $\nu$ represents the traded number of stocks per period.

Given that the number of stocks held at time $t$ is $\frac{\omega(t)W(t)}{S(t)}$, we derive an explicit expression for $\nu$ by applying It$\hat{\text{o}}$'s lemma to $\nu=\delta\big(\frac{\omega W}{S}\big)$ and keeping terms of order $O(\sqrt{\delta t})$:
\begin{equation*}
\nu=\frac{(\omega-1)\omega}{S}\cdot\bigg[\frac{\beta LW \delta B^\gamma+\sqrt{v}W\delta B^S}{1+\kappa_{TC}\cdot \text{sign}(\nu)\cdot \omega}\bigg].
\end{equation*}
While the precise number of traded stocks cannot be determined in advance, we can calculate the expected transaction costs in a time step as follows:
\begin{equation}
\mathbb{E} \{ \kappa_{TC} S |\nu| \}=\frac{\kappa_{TC}}{1+\kappa_{TC}\cdot \text{sign}(\nu)\omega}\cdot |\omega-1|\omega W\cdot \mathbb{E}\bigg\{ \bigg| \beta L \delta B^\gamma+\sqrt{v} \delta B^S \bigg| \bigg\}.
\label{E_init}
\end{equation}
Noting the transaction costs rate $\kappa_{TC}<1$ and the fraction $\omega \in [0, 1]$, $\kappa_{TC}^2 \omega \ll 1$, then
\begin{equation}
\frac{\kappa_{TC}}{1+\kappa_{TC} \cdot \text{sign}(\nu)\omega}\approx\kappa_{TC} \pm \kappa_{TC}^2 \omega \approx \kappa_{TC}.
\end{equation}
To compute the expected absolute value of the sum of two correlated Brownian motions in Eq. (\ref{E_init}), we firstly re-write $\delta B^\gamma$ and $\delta B^S$ as
\begin{align*}
\begin{cases}
\delta B^\gamma &=\sqrt{\delta t}Z_1,\\
\delta B^S &=\rho_4\sqrt{\delta t}Z_1+\sqrt{1-\rho_4^2}\sqrt{\delta t}Z_2,
\end{cases}
\end{align*}
where $Z_1, Z_2 \sim \mathscr{N}(0, 1)$, and thus
\begin{equation*}
\mathbb{E}\bigg\{ \bigg| \beta L \delta B^\gamma+\sqrt{v} \delta B^S \bigg| \bigg\}=\sqrt{\frac{2}{\pi}}\cdot\sqrt{(\beta L+\rho_4\sqrt{v})^2+(1-\rho_4^2)v}\cdot  \sqrt{\delta t}
\end{equation*}
Then the expected proportional transaction costs in one time step can be approximated as
\begin{equation*}
\mathbb{E} \{ \kappa_{TC} S |\nu| \}=\sqrt{\frac{2}{\pi \delta t}}\kappa_{TC} (1-\omega)\omega W \sqrt{(\beta L+ \rho_4 \sqrt{v})^2+(1-\rho_4^2)v}\cdot \delta t. 
\end{equation*}
Therefore, in an illiquid market with proportional transaction costs under two-factor stochastic volatility, the investor's wealth process follows the following dynamics:
\begin{equation}
\label{dynamics-wealth}
\delta W=[r W + (\mu -r)\omega W]\delta t+\beta\omega L W \delta B^\gamma +\omega\sqrt{v}  W \delta B^S-\sqrt{\frac{2}{\pi \delta t}}\kappa_{TC} (1-\omega)\omega W \sqrt{(\beta L+ \rho_4 \sqrt{v})^2+(1-\rho_4^2)v}\cdot \delta t.
\end{equation}

We now formulate a utility maximization model for an investor allocating his or her wealth dynamically between bonds and stocks. The investor's objective is to maximize the expected utility of terminal wealth at time $T$ by employing admissible trading strategies $\mathscr{A}$, which adjust portfolio weights between these two assets. The corresponding value function $Q$ is defined as
\begin{equation}
Q(W, v, \theta, L, t)=\max_{\omega \in \mathscr{A}} \mathbb{E}_t \bigg\{ U(W_T) \bigg| W_t=W, v_t=v, \theta_t=\theta, L_t=L \bigg\},
\end{equation}
where $U(\cdot)$ denotes the investor's utility function. The choice of $U(\cdot)$ is crucial as it should accurately reflect the investor's risk aversion. Rather than adopting classic utility functions (e.g., exponential, logarithmic or power utilities) directly, we derive the utility function empirically from option prices. Our approach combines statistical methods with machine learning techniques, achieving substantially better performance compared to single classic utility specifications.

With the dynamics of state variables specified in Eqs. (\ref{dynamics-stock}) and (\ref{dynamics-wealth}), the HJB equation is derived as
\begin{equation}
\label{Q}
\max_{\omega \in [0, 1]} \bigg\{\mathscr{L} Q(W, v, \theta, L, t) \bigg\}=0, \quad \forall (W, v, \theta, L, t) \in \Omega_T,
\end{equation}
where $\Omega_T = \R_+ \times \R_+  \times \R_+ \times \R_+\times [0,T]$, and the operator $\mathscr{L}$ is given by
\begin{align}
\label{L}
\mathscr{L} &=
\frac{\partial }{\partial t}+\bigg(r + (\mu -r)\omega -\sqrt{\frac{2}{\pi \delta t}}\kappa_{TC} (1-\omega)\omega \sqrt{(\beta L+\rho_4\sqrt{v})^2+(1-\rho_4^2)v}\bigg)W\frac{\partial }{\partial W}\\ \nonumber
&+\frac{1}{2}\bigg(\beta^2 L^2+v+2\rho_4\beta \sqrt{v} L\bigg)\omega^2 W^2 \frac{\partial ^2 }{\partial W^2}+\kappa(\theta-v)\frac{\partial }{\partial v}+\frac{1}{2}\sigma_1^2v\frac{\partial ^2}{\partial v^2}+\lambda(\eta-\theta)\frac{\partial }{\partial \theta}\\\nonumber
&+\frac{1}{2}\sigma^2_2\theta\frac{\partial ^2}{\partial \theta^2}+\alpha(\hat{\theta}+\lambda_{TC} \kappa_{TC} L^{\xi}-L) \frac{\partial }{\partial L}+\frac{1}{2}\sigma_L^2 \frac{\partial ^2 }{\partial L^2}+\rho_1\sigma_1 v \omega W\frac{\partial ^2}{\partial W\partial v}\\\nonumber
&+\rho_2\sigma_2\sqrt{v\theta}\omega W \frac{\partial ^2}{\partial W\partial \theta}+\bigg(\rho_6 \beta L+\rho_5 \sqrt{v}\bigg)\sigma_L \omega W \frac{\partial ^2 }{\partial W\partial L}+\rho_3\sigma_1\sigma_2\sqrt{v\theta}\frac{\partial^2}{\partial v \partial \theta}, \nonumber
\end{align}
with the terminal condition $Q(W, v, \theta, L, T)=U(W)$.

\section{The option-implied utility function}
The utility function plays a fundamental role in portfolio optimization models by providing a quantitative characterization of investor risk aversion and formally establishing the risk-return tradeoff. If the utility function is improperly specified, the portfolio optimization framework may become invalid. To develop an appropriate utility specification, we derive the utility function empirically from option prices \citep{Kang2006} by exploiting the theoretical relationship between: (i) the subjective probability density function (PDF) $P$, (ii) the risk-neutral probability density function (RN-PDF) $Q$, and (iii) the utility function itself:
\begin{equation}
P(S_T)=\frac{Q(S_T)/U'(S_T)}{\int \big(Q(x)/U'(x)\big)dx}.
\end{equation}

Then, once the estimates of $P$ (subjective PDF) and $Q$ (RN-PDF) are obtained, we can select a well-behaved functional form for the utility function and calibrate its parameters using machine learning techniques.
\subsection{Data}
China's options market has grown rapidly in recent years, supported by one of the world's largest retail and institutional investor bases. To capture utility-based investor preferences in such an active market, we analyze CSI 300 ETF options traded on the Shanghai Stock Exchange (SSE) from June 2020 to June 2024. These European-style options expire on the fourth Wednesday of each month. In this study, we use SSE-reported settlement prices and derive the risk-free rate from Shanghai Interbank Offered Rate (Shibor) overnight rates.

The Chinese options market exhibits distinctive characteristics that necessitate specialized data processing. Our study focuses on pronounced liquidity clustering around at-the-money (ATM) options, with severe illiquidity in deep in-the-money (ITM) or out-of-the-money (OTM) contracts. To address this liquidity concentration while preserving more valuable option types, we filter out contracts with daily trading volumes below 10,000 - a threshold where delta values approach 0 or 1 (indicating deep ITM or deep OTM positions, respectively) - rather than eliminating all ITM options.

Following the aforementioned processing steps, we eliminate options that violate general arbitrage constraints, exhibit implied volatility exceeding 100\%, or are priced below two minimum tick sizes. Ultimately, we retain expiration series containing at least five valid option contracts for subsequent analysis.

\subsection{Estimation of the risk-neutral probability density function}
Following the results of \cite{breeden1978prices}, we estimate the RN-PDF through second-order differentiation of option prices:
\begin{equation}
q(S_T) = e^{r t} \left. \frac{\partial^2 C(K, t)}{\partial K^2} \right|_{K = S_T},
\end{equation}
where $C(K, t)$ denotes the price of a European call option at time $t$ with strike price $K$. 

To ensure accurate conversion from implied volatilities to call option prices, we employ the smoothed implied volatility smile method \citep{shimko1993bounds} to obtain fitted implied volatilities. Specifically, we apply the cubic spline method \citep{bliss2002testing} on the option’s delta spaces to guarantee the fitted volatility curve satisfies arbitrage-free conditions. For extrapolation beyond observable strike prices, we extend the spline function horizontally outside the data range following the idea of \cite{bliss2004option}. Then the parameter estimation can be formalized as the following optimization problem:
\begin{equation}
\min_{\phi} \left[ \lambda \sum_{i=1}^n w_i \bigl[ y_i - f(x_i; \phi) \bigr]^2 + (1 - \lambda) \int \bigl[ f''(x; \phi) \bigr]^2 \, dx \right],
\end{equation}
where $x_i$ and $y_i$ denote the delta and implied volatility of option $i$ respectively, $f(x; \phi)$ represents the fitted spline function with parameter matrix $\phi$, $w_i$ is the weighting factor for observation $i$, determined by its proportional daily trading volume. The smoothing parameter $\lambda$  is set to $0.99$ to ensure robust fitting performance \citep{bliss2004option}.

Estimating the RN-PDF reduces to solving an optimization problem efficiently, for which we employ the quasi-Newton algorithm—a second-order method in machine learning. Compared to first-order methods like gradient descent, this approach converges significantly faster while avoiding expensive second-order derivative computations. It demonstrates particular advantages for large-scale parameter optimization, exhibiting numerical stability, robustness, and insensitivity to initial point selection. These characteristics make it especially suitable for our problem of large parameter matrix estimation.

\subsection{Testing the forecast ability of derived subjective PDFs}
Once the RN-PDFs are obtained, one needs to test whether the subjective PDFs estimated via RN-PDFs and utility functions demonstrate strong forecast ability. We first establish the null hypothesis that the estimated subjective PDFs are valid and that option payoffs across different expiration dates are independent. Under this null hypothesis, the inverse probability transformations of the realizations should be independent and uniformly distributed, as specified below:
\begin{equation}
y_t = \int_{-\infty}^{X_t} \hat{P}_t(s) \, \mathrm{d}s \sim \text{i.i.d.}\ \mathcal{U}(0,1),
\end{equation}
where $X_t$ is a realization at an option expiration date and $\hat{P}_t(\cdot)$ is an estimated subjective PDF.

Following the work of \cite{berkowitz2001testing}, we conduct a joint test for both uniformity and independence by estimating the following AR(1) model and performing a likelihood ratio test:
\begin{equation}
\label{Z_t}
Z_t - \mu = \rho (Z_{t-1} - \mu) + \varepsilon_t,
\end{equation}
where $Z_t = \Phi^{-1}(y_t)$, with $\Phi$ being the standard normal cumulative PDF.

Under the null hypothesis, the parameters in the above equation should satisfy: $\mu = 0, \rho = 0$, and $\sigma_{\varepsilon_t} = 1$. Then we estimate the utility function parameters  by solving the likelihood ratio minimization problem:
\begin{equation}
\min \mathrm{LR3} = -2 \left[ L(0,1,0) - L(\mu, \sigma^2, \rho) \right],
\end{equation}
using the quasi-Newton algorithm, where $L(\mu, \sigma^2, \rho)$ represents the log-likelihood function from Eq. (\ref{Z_t}).


We employ the LR3-statistic to jointly test for uniformity and independence, and the LR1-statistic (defined as $= -2 \left[ L(\hat{\mu}, \hat{\sigma}^2, 0) - L(\hat{\mu}, \hat{\sigma}^2, \hat{\rho}) \right]$) to test independence individually. Through analysis of these test statistics, we assess whether the estimated subjective PDFs match the true PDFs. When both statistics fail to reject the null hypotheses, we conclude that the derived subjective PDFs accurately predict the true PDFs.


\subsection{Selecting utility functions using machine learning}
We now select an appropriate utility function form whose generated subjective PDFs demonstrate strong forecasting accuracy for the true PDFs. In addition to classical utility functions adopted in \cite{bliss2004option, Kang2006} (see Appendix \ref{Utilityfct} for specifications), we incorporate an S-shaped utility function from prospect theory. The S-shaped function captures loss aversion psychology, characterized by concave utility in the gain domain and convex utility in the loss domain, which makes it particularly well-suited for modeling investor behavior.

However, classical S-shaped utility functions adopted in \cite{kai1979prospect, hu2023casino} lack smooth differentiability at the reference point, making them unsuitable for our analysis, requiring continuously differentiable functions. To overcome this limitation, we therefore adopt the hyperbolic tangent utility function recently proposed by \cite{adjei2025prospect}, which preserves prospect-theoretic curvature while guaranteeing continuous differentiability. This formulation offers dual advantages: direct parameter control for smoothness requirements and inherited desirable nonlinear properties from its widespread use as a machine learning activation function. The explicit functional form of our selected S-shaped utility function is as follows:
\begin{equation}\label{S-shaped utility}
U(W) = \begin{dcases}
      \tanh(k_1(W - W_{0})), & \text{if } W \geq W_{0}\\
       -\frac{k_1}{k_2} \tanh(k_2(W_{0} - W)), & \text{if } W < W_{0}
\end{dcases},
\end{equation}
where $W_{0}$ denotes the reference point that endogenously partitions outcomes into gain and loss domains, with ($k_1$, $k_2$) parameterizing the differential risk attitudes in these two domains.

To evaluate the forecasting performance of the subjective PDFs generated by these utility functions, we compare their Berkowitz p-values across four distinct forecast horizons, as presented in Table \ref{tab:pdf_stats}. All reported p-values are adjusted using Monte Carlo tests as proposed by \cite{bliss2004option}, where actual realizations of underlying asset prices are repeatedly replaced with pseudo-realizations.
\begin{table}[H]
\centering
\small
\caption{Berkowitz statistic p-values(adjusted)}
\label{tab:pdf_stats}
\begin{tabularx}{\textwidth}{@{}l *{3}{>{\centering\arraybackslash}X}@{}}
\toprule
\textbf{Forecast Horizon} & 
\textbf{PDF} & 
\textbf{LR3 p-value} & 
\textbf{LR1 p-value} \\
\midrule
\multirow{7}{*}{1 weeks}  
& Risk neutral       & 0.007 & 0.150 \\
& Power              & 0.057 & 0.098 \\
& Exponential        & 0.076 & 0.087 \\
& HARA               & 0.004 & 0.154 \\
& Log-power          & 0.053 & 0.099 \\
& Linear-exponential & 0.073 & 0.087 \\
& S-type             & 0.211 & 0.085 \\
\cmidrule(lr){2-4}

\multirow{7}{*}{2 weeks}
& Risk neutral       &0.033 & 0.158 \\
& Power              &  0.101 & 0.117 \\
& Exponential        & 0.099 & 0.091 \\
& HARA               & 0.017 & 0.162 \\
& Log-power          & 0.101 & 0.091 \\
& Linear-exponential &0.098  & 0.117 \\
& S-type             & 0.199 & 0.106 \\
\cmidrule(lr){2-4}

\multirow{7}{*}{3 weeks}
& Risk neutral       & 0.027 & 0.038 \\
& Power              & 0.048 & 0.039 \\
& Exponential        & 0.049 & 0.040 \\
& HARA               & 0.014 & 0.038 \\
& Log-power          & 0.048 & 0.039 \\
& Linear-exponential & 0.049 & 0.040 \\
& S-type             & 0.033 & 0.043 \\
\cmidrule(lr){2-4}

\multirow{7}{*}{4 weeks}
& Risk neutral       & 0.000 & 0.000 \\
& Power              & 0.000 & 0.000 \\
& Exponential        & 0.000 & 0.000 \\
& HARA               &0.000 & 0.000 \\
& Log-power          & 0.000 & 0.000 \\
& Linear-exponential & 0.000 & 0.000 \\
& S-type             &0.001 & 0.000 \\
\bottomrule
\end{tabularx}
\end{table}


As shown in the Table above, at the three- and four-week horizons, all utility functions fail the LR1 independence test. For the remaining two horizons, most utility functions pass the tests, with the S-shaped specification showing statistically superior performance across all model specifications. These results confirm that the S-shaped utility function generates subjective PDFs with significantly greater predictive capability than alternative functional forms. Based on a comprehensive horizon analysis, we selected the parameter-optimized S-shaped utility function (Eq. \ref{S-shaped utility}) at the two-week forecast horizon for subsequent calculations, with the parameters $k_1=2.27$, $k_2=2.81$, and $W_0=4.76$ determined via quasi-Newton optimization.


\subsection{Concavification of the S-shaped utility}
Considering S-shaped utility in a portfolio optimization problem introduces significant mathematical complexities that require careful treatment. The fundamental challenge arises from the fact that the S-shaped utility function destroys the concavity of the value function, causing the associated HJB equation to lack a unique classical solution or even a well-defined viscosity solution in certain regions of the state space. Following the comparison principle established by \cite{dai2022nonconcave}, under their new definition of viscosity solution, we replace the S-shaped utility with its concave envelope, thereby reducing the original non-concave utility maximization problem to a concave one. This reformulated problem can then be solved using standard analytical or numerical approaches.

\begin{figure}[htbp]
  \centering
  \includegraphics[width=0.8\textwidth]{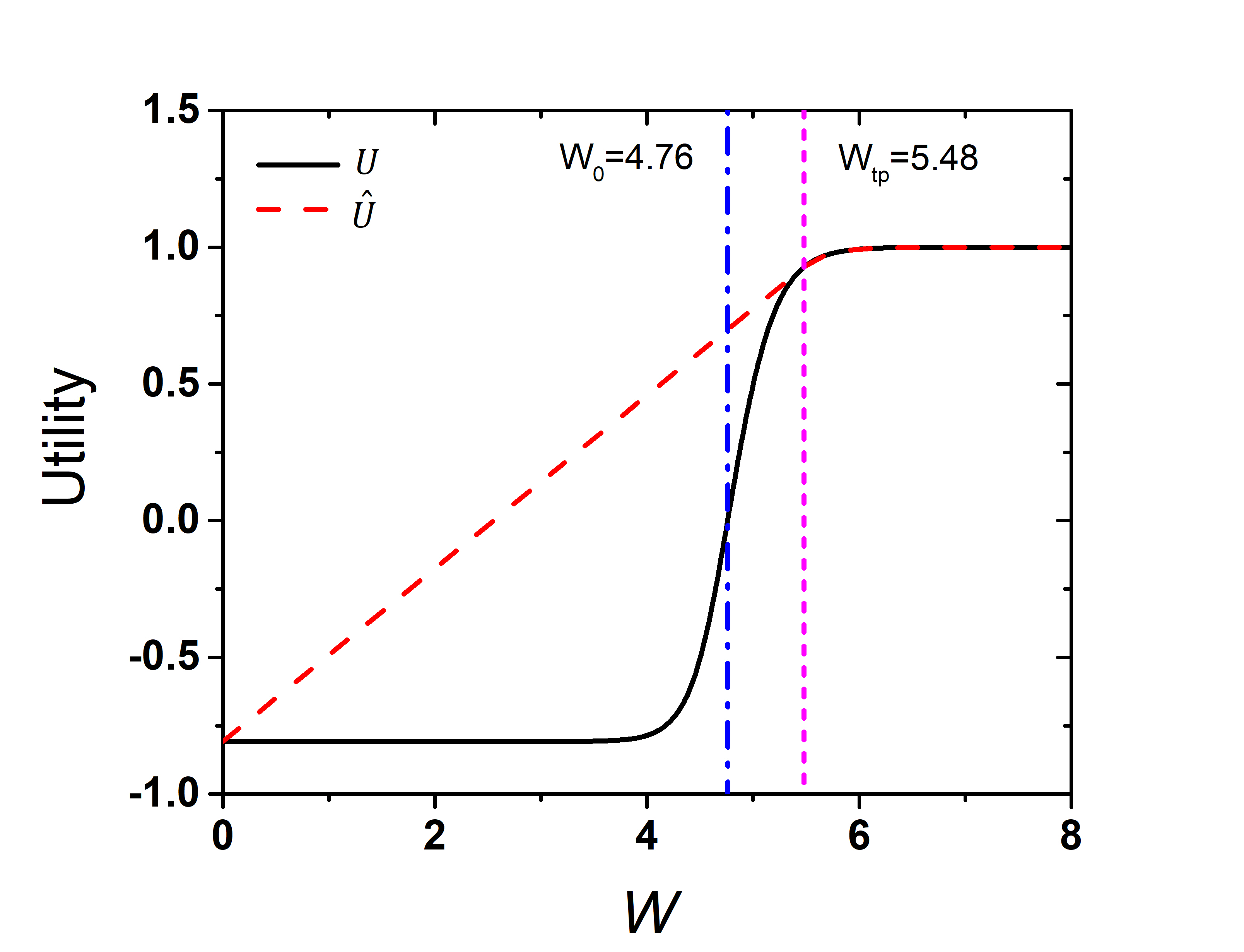}  
  \caption{The original S-shaped utility and its concave envelope}
  \label{fig:demo}  
\end{figure}

For clarity, let $U_1(W)$ and $-U_2(-W)$ denote the first and second functions of the S-shaped utility function in Eq. (\ref{S-shaped utility}), respectively. Then, there exists a function $g\colon [0,\infty) \to [0,\infty)$ such that $g(W_0) \geq W_0$ for all $W_0 \in [0,\infty)$, satisfying:
\begin{equation}
U_{1}(g(W_0)-W_0)+U_{2}(W_0)-g(W_0) U_{1}^{\prime}(g(W_0)-W_0)=0,
\end{equation}
and the concave envelope is given by:
\begin{equation}
\hat{U}(W) = 
\begin{cases}
\tanh\left(2.27 \cdot (W - 4.76)\right), & W \geq 5.48\\
-0.81 + 0.32 \cdot W, & W < 5.48\\
\end{cases},
\label{eq:concave_envelope}
\end{equation}
with $W_{tp} \approx 5.48$ being the tangent point of the line from $(0, U(0))$ to the original utility function. The concave envelope $\hat{U}$ is constructed by combining this tangent line for lower wealth levels ($W<W_{tp}$) with the original concave segment of the utility function for higher wealth levels ($W\geq W_{tp}$). As a result, $\hat{U}$ is a concave, monotonically increasing, and $C^1$-smooth function. It should be noted that since the tangent line is linear in wealth, the investor exhibits risk-neutral behavior within this region. Consequently, for the case where $\mu > r$, the optimal allocation to the risky asset is 100\% when wealth is below $W_{tp}$. To illustrate the construction of the concave envelope, we plot the resulting function in Figure \ref{fig:demo}.

\section{Numerical experiments}

\subsection{Deep learning-driven policy iteration scheme}
We note that the operator in Eq. (\ref{L}) is quadratic in $\omega$, which allows the optimal policy for the corresponding maximization problem to be expressed explicitly under certain conditions. However, the quotient involving partial derivatives may cause computational issues. To efficiently find optimal policies for our high-dimensional portfolio selection problem, we avoid classical mesh-based iteration schemes, as they are computationally expensive and suffer from the curse of dimensionality. Instead, we employ the method derived from PINNs \citep{raissi2019physics}, a deep learning technique for solving the resulting PDEs. PINNs have emerged as a powerful technique in recent years, achieving remarkable success across a wide range of applications. In our algorithm, according to the universal approximation theorem \citep{funahashi1989}, both the value function and the optimal policy are approximated by corresponding well-constructed neural networks. Specifically, we define the value network $Q_{\phi}:\Omega_{T}\to\mathbb{R}$ and the policy network $\omega_{\psi}:\Omega_{T}\to[0,1]^{1}$ as three-layer neural networks:
\begin{align}\label{networks}
Q_{\phi}&=f_{3}^{Q} \circ\tanh \circ f_{2}^{Q}\circ\tanh \circ f_{1}^{Q},\\
\omega_{\psi}&=\operatorname{sigmoid}\circ f_{3}^{\omega}\circ\tanh\circ f_{2}^{\omega}\circ\tanh\circ f_{1}^{\omega},\nonumber
\end{align}
where $\phi$ and $\psi$ denote the network parameters, and the description of the network structures is detailed in Appendix \ref{Neutral work}. 

Subsequently, in the policy evaluation step, the associated PDE is solved for a given policy by approximating the value function $Q$ with a neural network. The solution is obtained by minimizing a loss function formed by a summation of the PDE residuals and the terminal condition residuals. Specifically, the loss function during the k-th iteration is defined as
\begin{equation}
\mathcal{L}^{(k)}(\phi) = 
    \mathbb{E}_{W,v,\theta,L,t} \left[ \left( \mathcal{L}^{\omega_{\psi^{(k-1)}}} Q_{\phi}(W,v,\theta,L,t) \right)^{2} \right] +
    \mathbb{E}_{W,v,\theta,L} \left[ \left| Q_{\phi}(W,v,\theta,L,T) - \hat{U}(W) \right|^{2} \right].
\end{equation}
In addition, to enhance the convergence behavior of value function, we adopt a strategic sampling approach that increases the density of interior collocation points relative to boundary points at a ratio of approximately 4:1. Following this, the value network parameters $\phi^{(k)}$ at the k-th iteration are determined, leading to the policy improvement step where the following optimization problem is solved:
\begin{equation}
\psi^{(k)} = \underset{\psi}{\operatorname{argmax}} \,
\mathbb{E}_{W,v,\theta,L,t} \left[ 
\mathscr{L}^{\omega_{\psi}} Q_{\phi^{(k)}} (W,v,\theta,L,t)
\right].
\end{equation}
This maximization step empirically implements the policy improvement procedure by refining the policy network approximation using the newly updated value network. As a result, through alternating steps of policy evaluation and improvement, we obtain the following iteration scheme for solving Eq. (\ref{Q}) with classical and concavified utility functions:
\begin{algorithm}[H]
\caption{The deep learning-driven policy iteration scheme}
\label{alg:policy_iteration}
\begin{algorithmic}[1]
\State Given initial values of trainable parameters of networks $\phi^{(0)},\psi^{(0)}$
\State Construct value network and control network
\For{$k=1,\ldots,N$}
    \State Conduct the step of policy evaluation by calculating
    \State $\phi^{(k)}=\underset{\phi}{\operatorname{argmin}}\mathbb{E}_{W,v,\theta,L,t}\left[\left(\mathcal{L}^{\omega_{\psi^{(k-1)}}}Q_{\phi}(W,v,\theta,L,t)\right)^{2}\right]+\mathbb{E}_{W,v,\theta,L}\left[\left|Q_{\phi}(W,v,\theta,L,T)-\hat{U}(W)\right|^{2}\right]$
    \If{the maximum relative difference between $Q_{\phi^{(k)}}$ and $Q_{\phi^{(k-1)}}$ is smaller than $10^{-4}$}
        \State \textbf{break}
    \EndIf
    \State Conduct the step of policy improvement by calculating
    \State $\psi^{(k)}=\underset{\psi}{\operatorname{argmax}}\mathbb{E}_{W,v,\theta,L,t}\left[\mathcal{L}^{\omega_{\psi}}Q_{\phi^{(k)}}(W,v,\theta,L,t)\right]$
\EndFor
\State \Return $Q_{\phi^{(k)}},\omega_{\psi^{(k)}}$
\end{algorithmic}
\end{algorithm}

\subsection{Validation and order of convergence of our numerical scheme}
Given the absence of an analytical solution for the HJB equation (\ref{Q}), we refer to the classical Merton's problem \citep{merton1975optimum} to validate our model formulation. This is done by adopting a power utility function $U(W) = \frac{W^{1-\gamma}}{1-\gamma}$ and configuring the parameters in Eq. (\ref{Q}) accordingly. The parameters are set as follows: $\gamma = 0.5$, $r = 0.02$, $T = 1$, $\mu = 0.05$, and the volatility $\sqrt{v}$ is fixed at $0.16$. All other parameters in the model are set to zero.

\begin{figure}[H]
  \centering
  \includegraphics[width=0.8\textwidth]{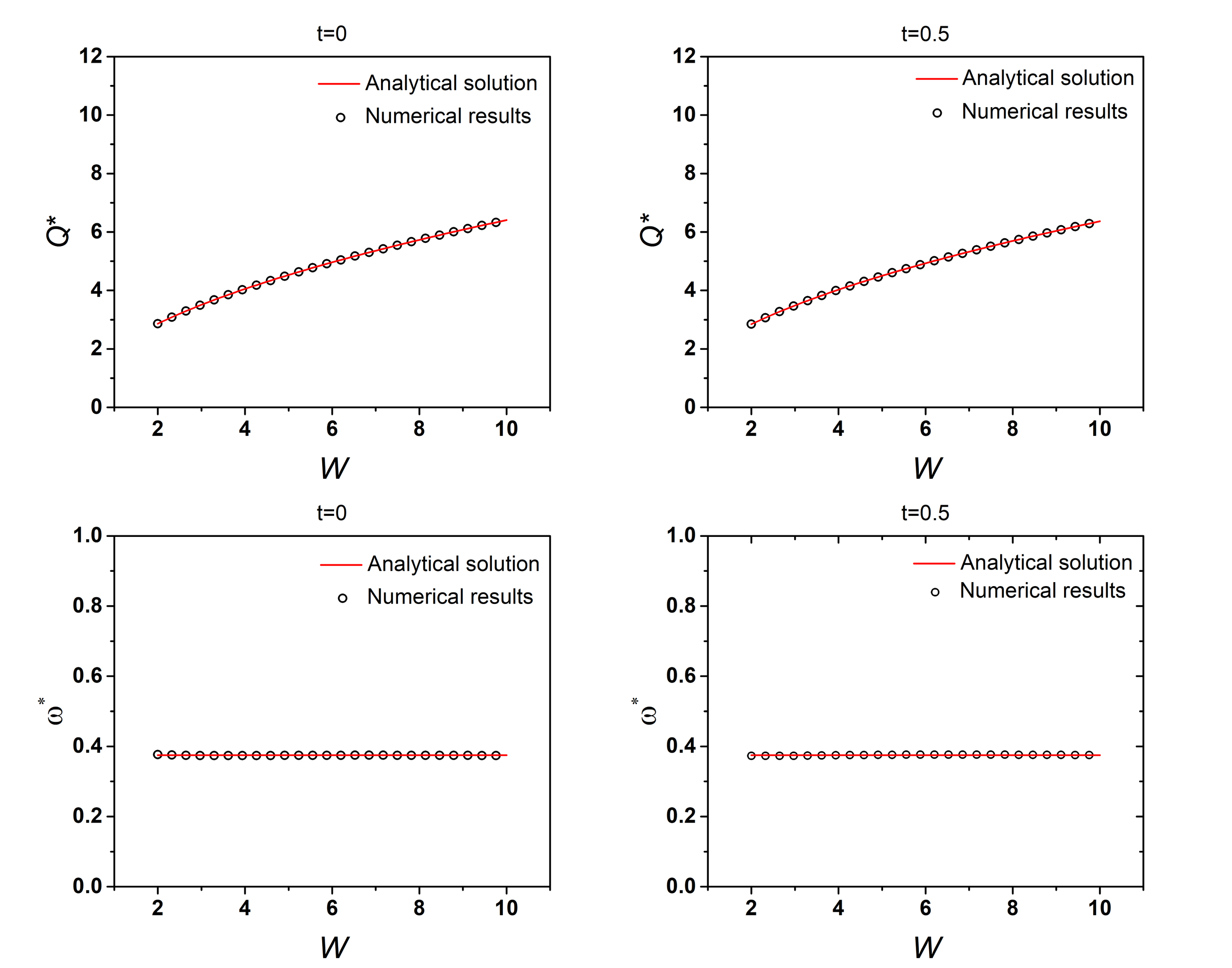}  
  \caption{Validation of numerical scheme with zero liquidity risk and zero transaction costs}
  \label{fig:2}  
\end{figure}

In Figure \ref{fig:2}, for both the values of the value function and the optimal policy, two time points are selected for analysis: t=0 and t=0.5, corresponding to the initial and intermediate stages of the strategy, respectively.  In this case, $\omega^{*}=\frac{\mu-r}{\gamma\sigma^{2}}=0.375$. The comparison shows that our algorithm yields a highly accurate estimation, closely matching the analytical solution.

\begin{figure}[H]
  \centering
  \includegraphics[width=0.8\textwidth]{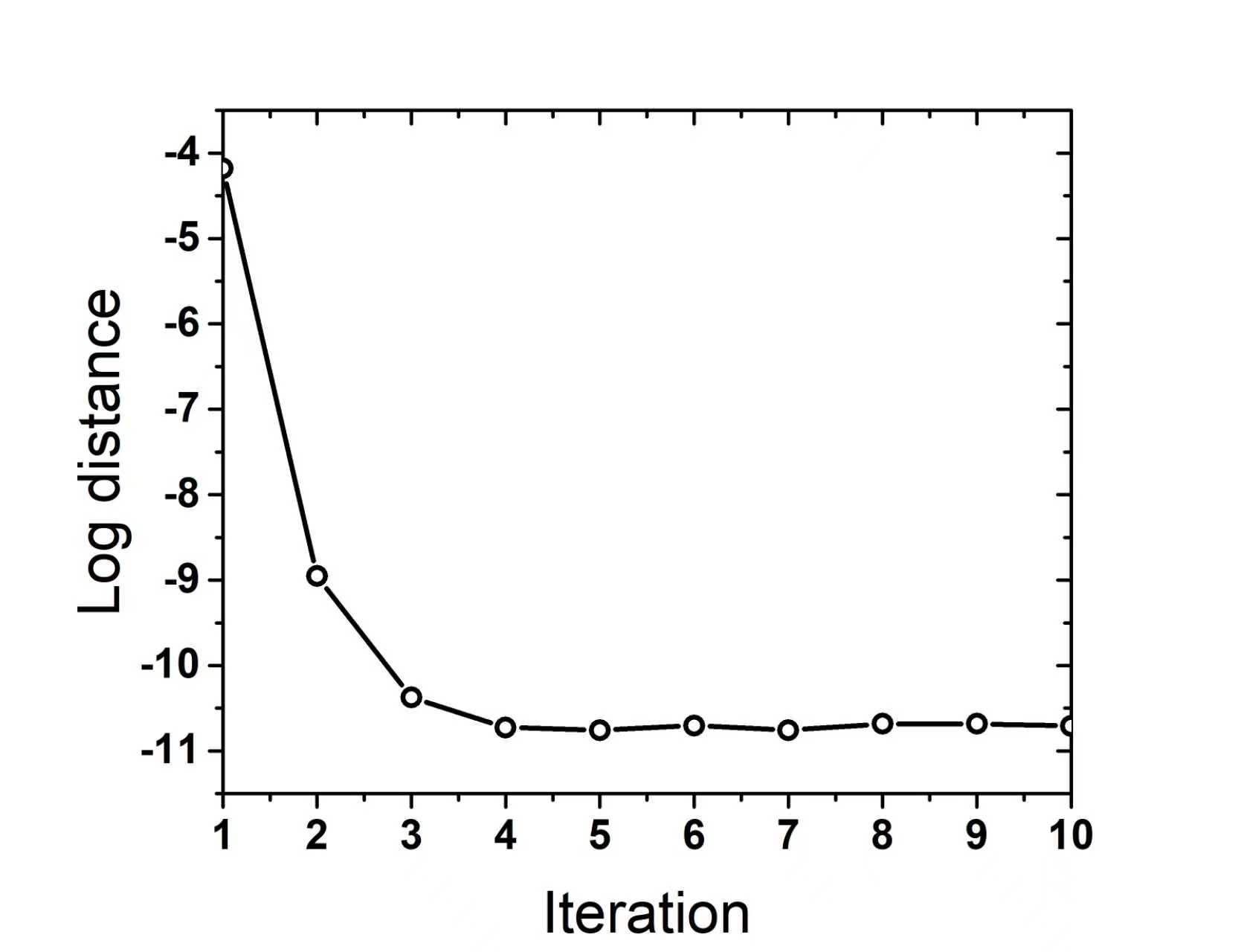}  
  \caption{Experimental convergence rates with zero liquidity risk and zero transaction costs}
  \label{fig:3}  
\end{figure}

The convergence behavior of our numerical scheme is illustrated in Figure \ref{fig:3}, which shows the log distance between our numerical results and the analytical solution to Merton's problem for each iteration. The rapid convergence of the value function after only a few iterations validates the robustness of our model within the Merton framework.

\subsection{Numerical experiments and discussions}
In this subsection, we return to the five-dimensional problem, aiming to investigate the convergence behavior of the value function and the numerical solution of the optimal policy under the assumption of an S-shaped utility function. Furthermore, by adjusting the parameters associated with exogenous and endogenous transaction costs, we examine how changes in these cost structures influence the optimal policy, i.e., the resulting shifts in the portfolio of the investors. 

Upon finalizing the form of the S-shaped utility function in the , this subsection examines how key parameters, i.e., those governing exogenous and endogenous transaction costs, as well as volatility, affect the optimal investment policy. We analyze the influence of variations in these parameters on the optimal strategy and evaluate whether the resulting changes are consistent with economic intuition. For clarity of exposition, we set $t = 0.5$ and fixed $W = 5.5$ generate two representative subplots illustrating how the optimal policy evolves with changes in other parameters. The remaining variables are fixed at $\theta = 0.2$, $v = 0.1$, $L = 0.3$, while all other parameter values are held at their baseline levels, as provided in Table \ref{tab:default_params}, unless otherwise specified.

\begin{table}[htbp]
\centering
\caption{Default Parameters}
\begin{tabular}{l >{$}c<{$} l >{$}c<{$}}
\toprule
\textbf{Parameter} & \textbf{Value} & \textbf{Parameter} & \textbf{Value} \\
\midrule
$r$      & 0.01        & $\bar{\theta}$ & 0.6        \\
$\mu$    & 0.05        & $\lambda$      & 1.5        \\
$\rho_1$ & 0.5         & $\alpha$       & 2.0        \\
$\rho_2$ & 0.2         & $\beta$        & 0.3        \\
$\rho_3$ & 0.3         & $\kappa$       & 5.0      \\
$\rho_4$ & 0.5         & $\lambda_{TC}$       &5.0      \\
$\rho_5$ & 0.5         & $\kappa_{TC}$       &0.4\%     \\
$\rho_6$ & 0.5         & $\sigma_1$       &0.1      \\
$\sigma_2$ & 0.1       & $\sigma_L$     & 0.2        \\
$\gamma$ & 0.5      & $\eta$       &0.15      \\
$\delta t$ & \frac{1}{12} & $T$          & 1          \\
\bottomrule
\label{tab:default_params}
\end{tabular}
\end{table}

\subsubsection{The changing of $\beta$}

\begin{figure}[h!]
  \centering
  \subfigure[The variation with wealth.]{\label{fig:beta-W}\includegraphics[width=130mm]{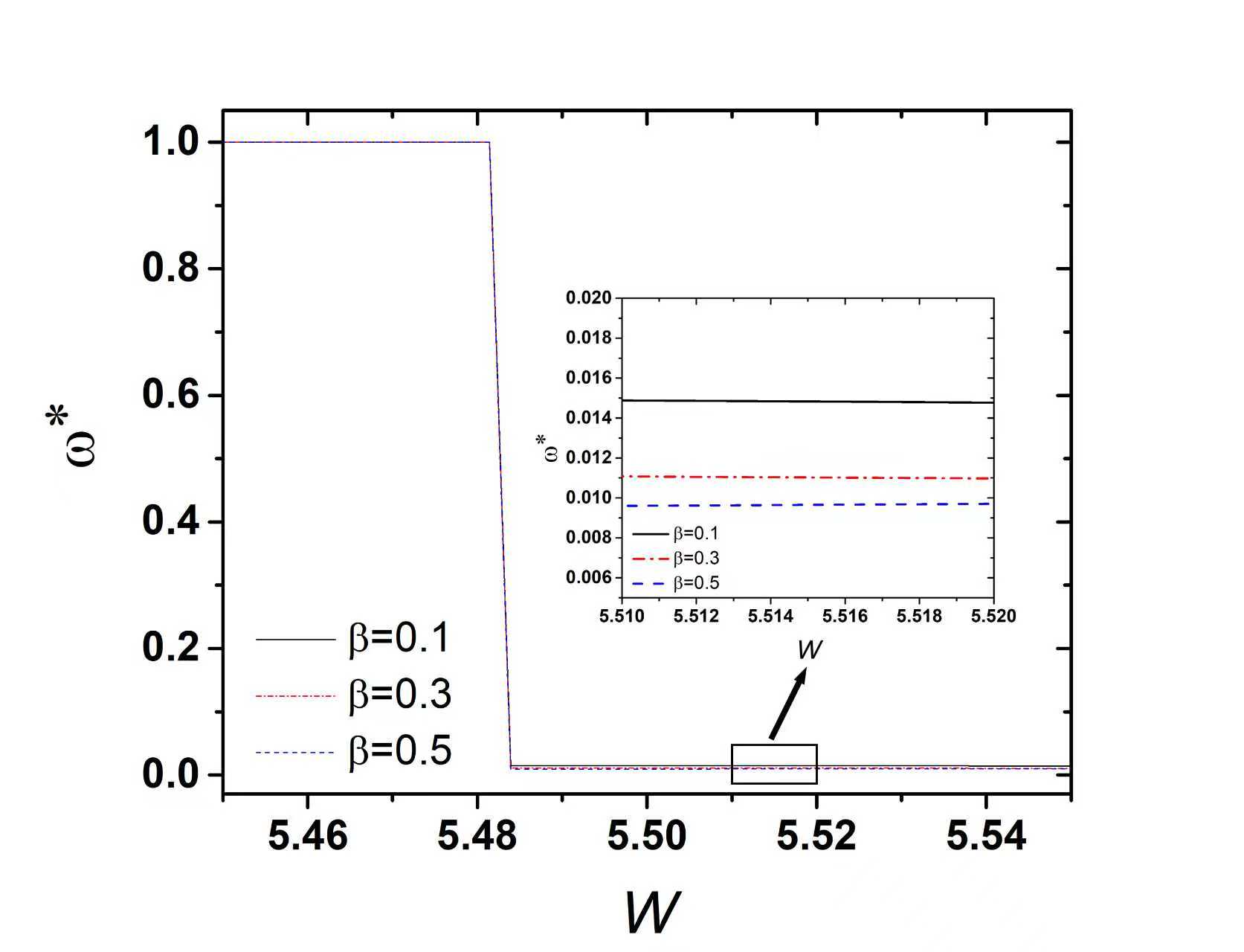}}
  \subfigure[The variation with time.]{\label{fig:beta-t}\includegraphics[width=130mm]{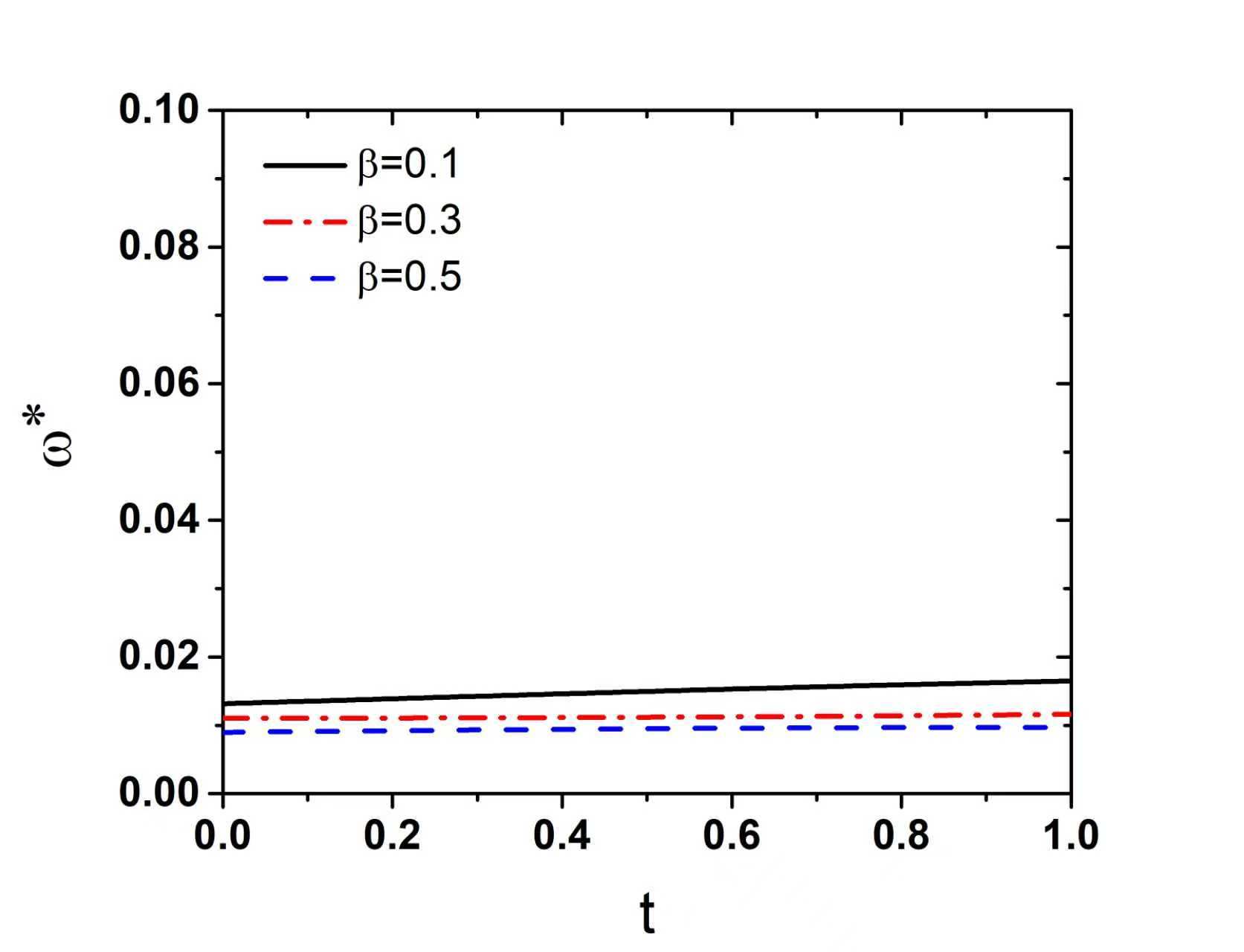}}
  \caption{Different $\beta$.} \label{fig:beta}
\end{figure}

We begin by analyzing the effect of $\beta$, a parameter that captures the sensitivity of the asset price to the level of market liquidity. Figure \ref{fig:beta} presents numerical results for $\beta = 0.1$, $0.3$, and $0.5$, with all other parameters held constant. At a wealth level of $W = 5$, situated in the convex (risk-averse) region of the S-shaped utility function, the investor exhibits cautious behavior. As $\beta$ increases, indicating that the asset price becomes more responsive to liquidity conditions, the optimal stock holding ratio declines. This response aligns with economic intuition: when asset values are more vulnerable to liquidity shocks, a risk-averse investor will reduce equity exposure to mitigate potential losses. Moreover, at a fixed time $t = 0.5$, the relationship between wealth and the optimal stock holding ratio exhibits an inverted S-shaped pattern. In the risk-averse region, a lower $\beta$, reflecting lower liquidity-driven price sensitivity, is associated with a higher optimal allocation to stocks. In contrast, within the risk-seeking region of the utility function, the pattern reverses: a higher $\beta$ leads to a greater stock holding ratio. This asymmetry underscores how investor response to liquidity risk depends critically on the underlying risk preference regime.

\subsubsection{The changing of $\kappa_{TC}$}

\begin{figure}[h!]
  \centering
  \subfigure[The variation with wealth.]{\label{fig:ktc-W}\includegraphics[width=130mm]{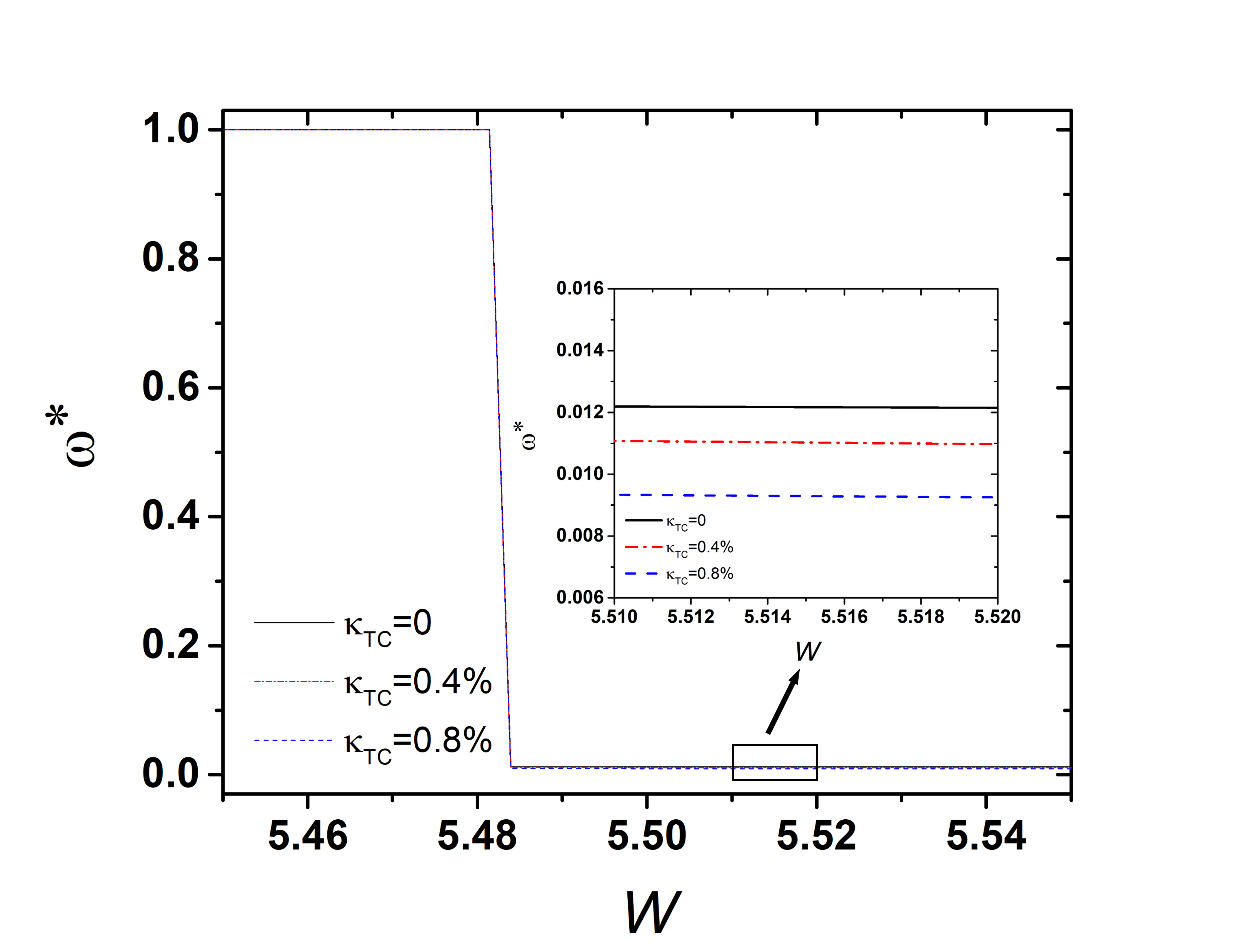}}
  \subfigure[The variation with time.]{\label{fig:ktc-t}\includegraphics[width=130mm]{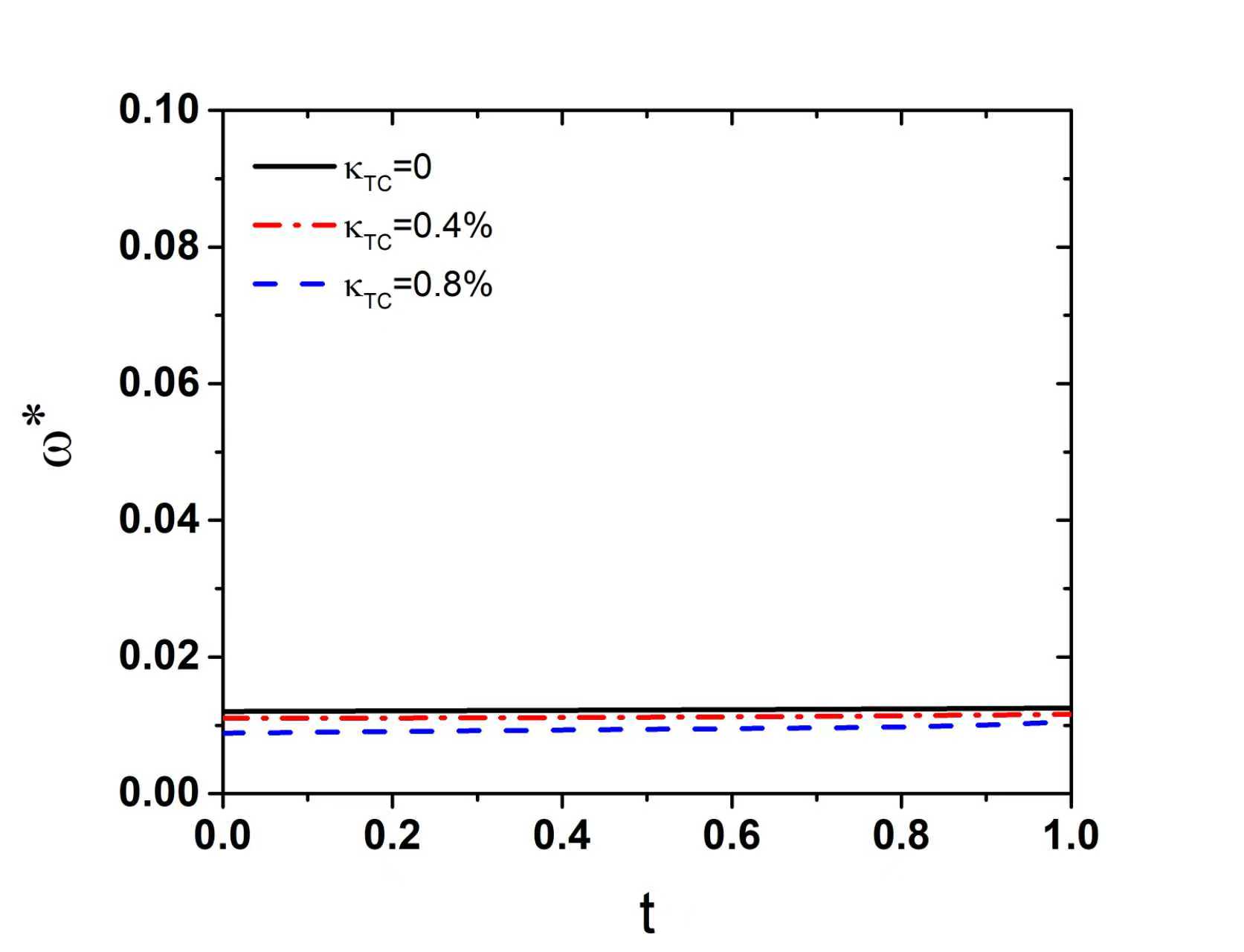}}
  \caption{Different $\kappa_{TC}$.} \label{fig:ktc}
\end{figure}

We now turn to the parameter $\kappa_{TC}$, which represents the proportional transaction cost rate. Figure \ref{fig:ktc} illustrates the optimal stock holding ratio for values of $\kappa_{TC} = 0.0\%$, $0.4\%$, and $0.8\%$, with all other parameters held constant. The results indicate that variation in $\kappa_{TC}$ influences the optimal strategy in a manner qualitatively similar to the liquidity sensitivity parameter $\beta$. Although the quantitative impact of $\kappa_{TC}$ is more moderate compared to that of $\beta$, as reflected by the narrower spread between the curves. Despite this weaker magnitude, the corresponding trend remains statistically and economically discernible, underscoring the importance of incorporating even modest frictions into portfolio choice models.

\subsubsection{The changing of $\sigma_L$}

\begin{figure}[H]
  \centering
  \subfigure[The variation with wealth.]{\label{fig:sigmaL-W}\includegraphics[width=130mm]{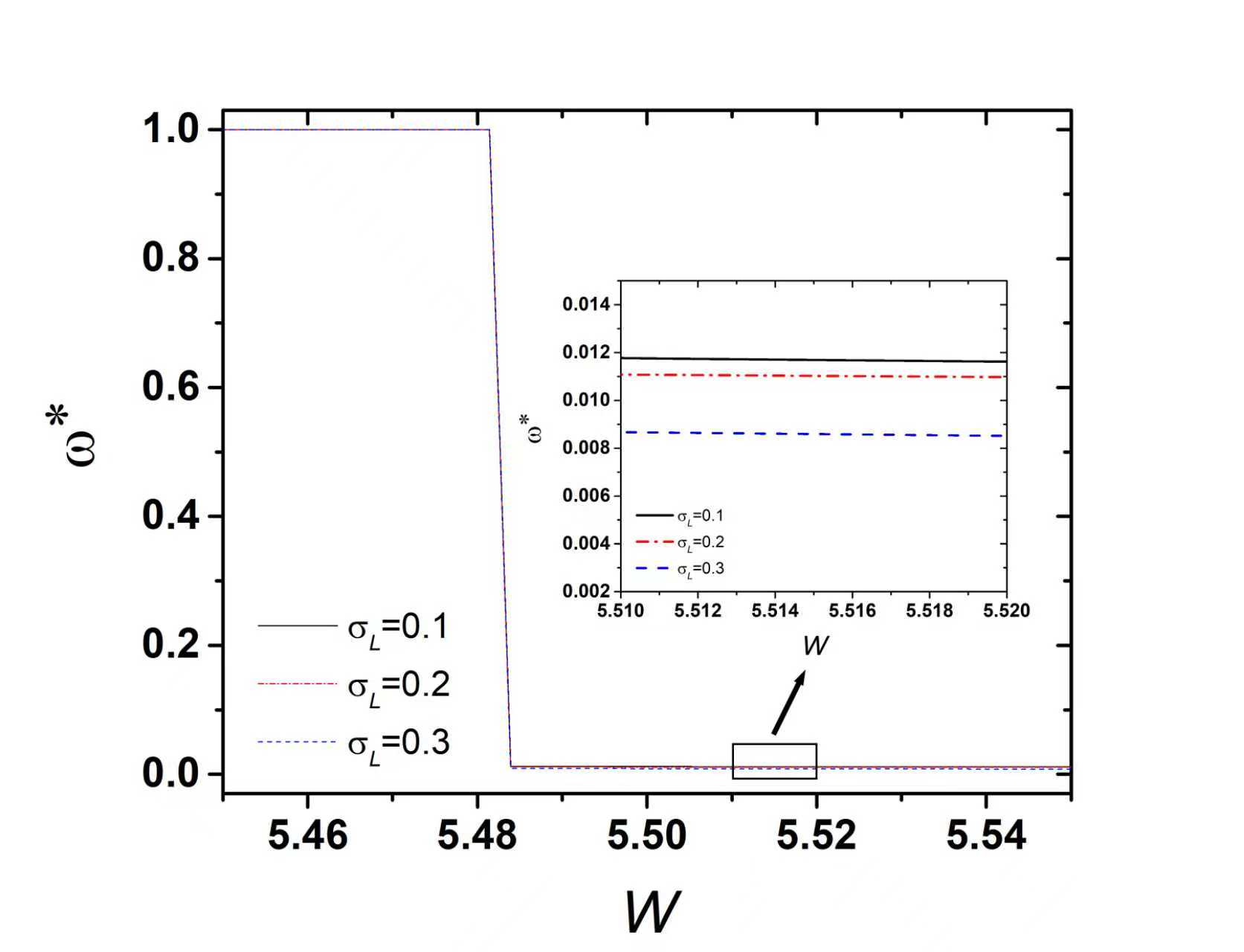}}
  \subfigure[The variation with time.]{\label{fig:sigmaL-t}\includegraphics[width=130mm]{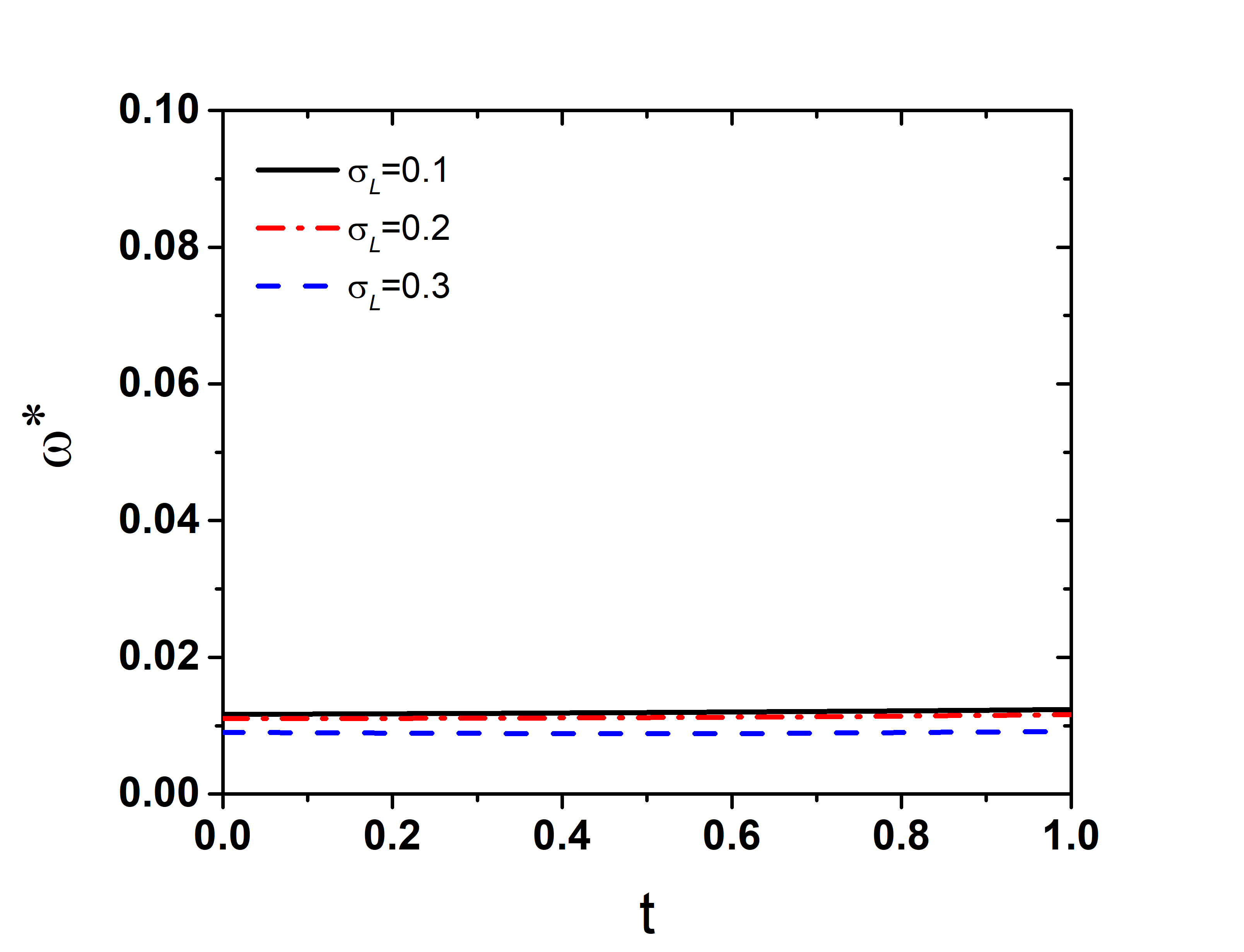}}
  \caption{Different $\sigma_{L}$.} \label{fig:sigmaL}
\end{figure}

We now examine the effect of $\sigma_L$, the volatility of market liquidity. Figure \ref{fig:sigmaL} displays the optimal stock proportion for values of $\sigma_L = 0.1$, $0.2$, and $0.3$, with all other parameters held constant. The results reveal that an increase in $\sigma_L$, while moderate in magnitude, leads to a discernible reduction in the optimal allocation to stocks. This outcome aligns with financial theory: heightened volatility in market liquidity amplifies the uncertainty of transaction execution and potential price impact, thereby increasing the implicit cost of trading. For a risk-averse investor, this elevated uncertainty serves as a deterrent to holding risky assets, prompting a shift toward a more conservative portfolio. The observed inverse relationship underscores the role of liquidity stability as a distinct risk factor in investment decisions, even when its quantitative effect is less pronounced than other market parameters.

\subsubsection{The changing of $v_0$}

\begin{figure}[H]
  \centering
  \subfigure[The variation with wealth.]{\label{fig:v0-W}\includegraphics[width=130mm]{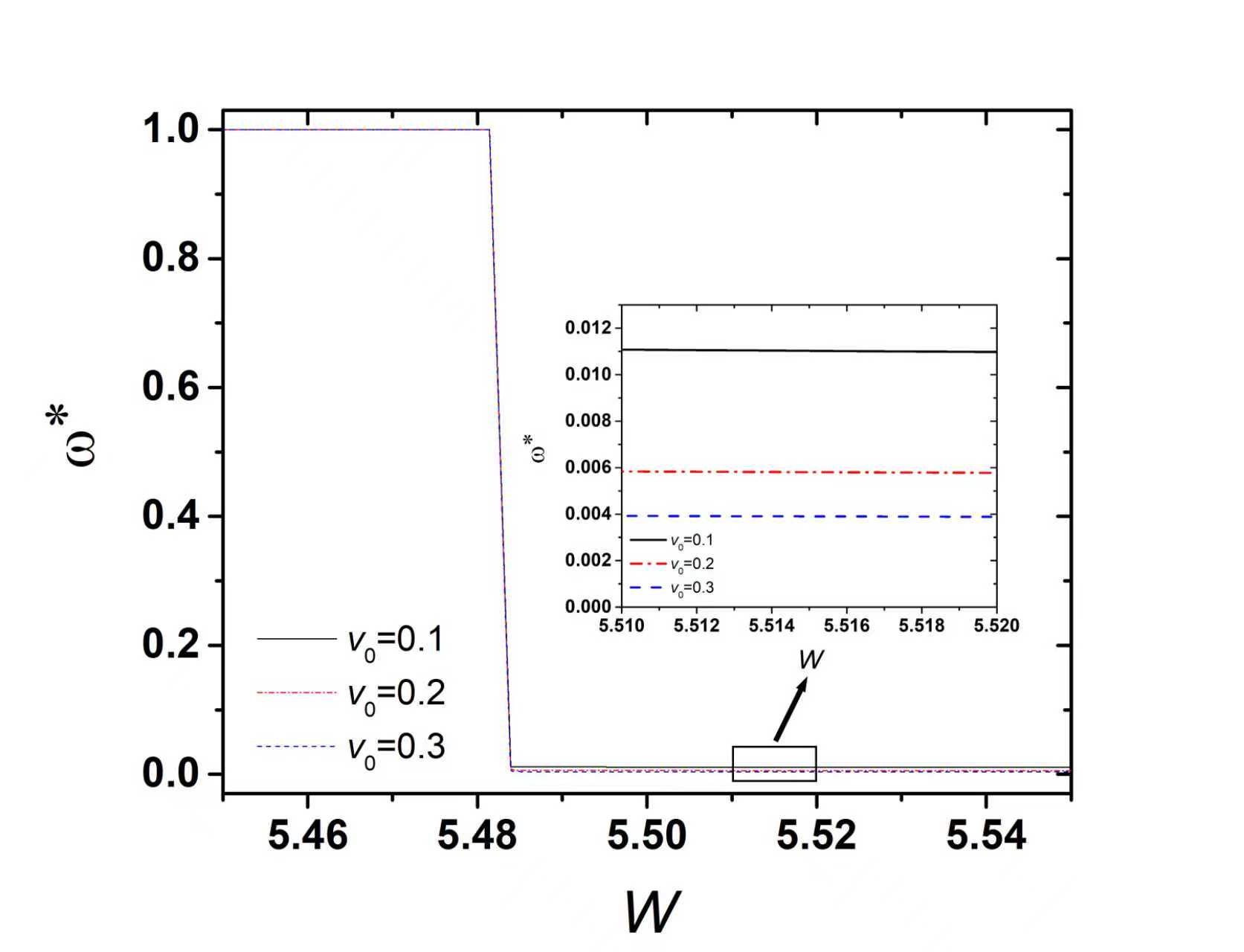}}
  \subfigure[The variation with time.]{\label{fig:v0-t}\includegraphics[width=130mm]{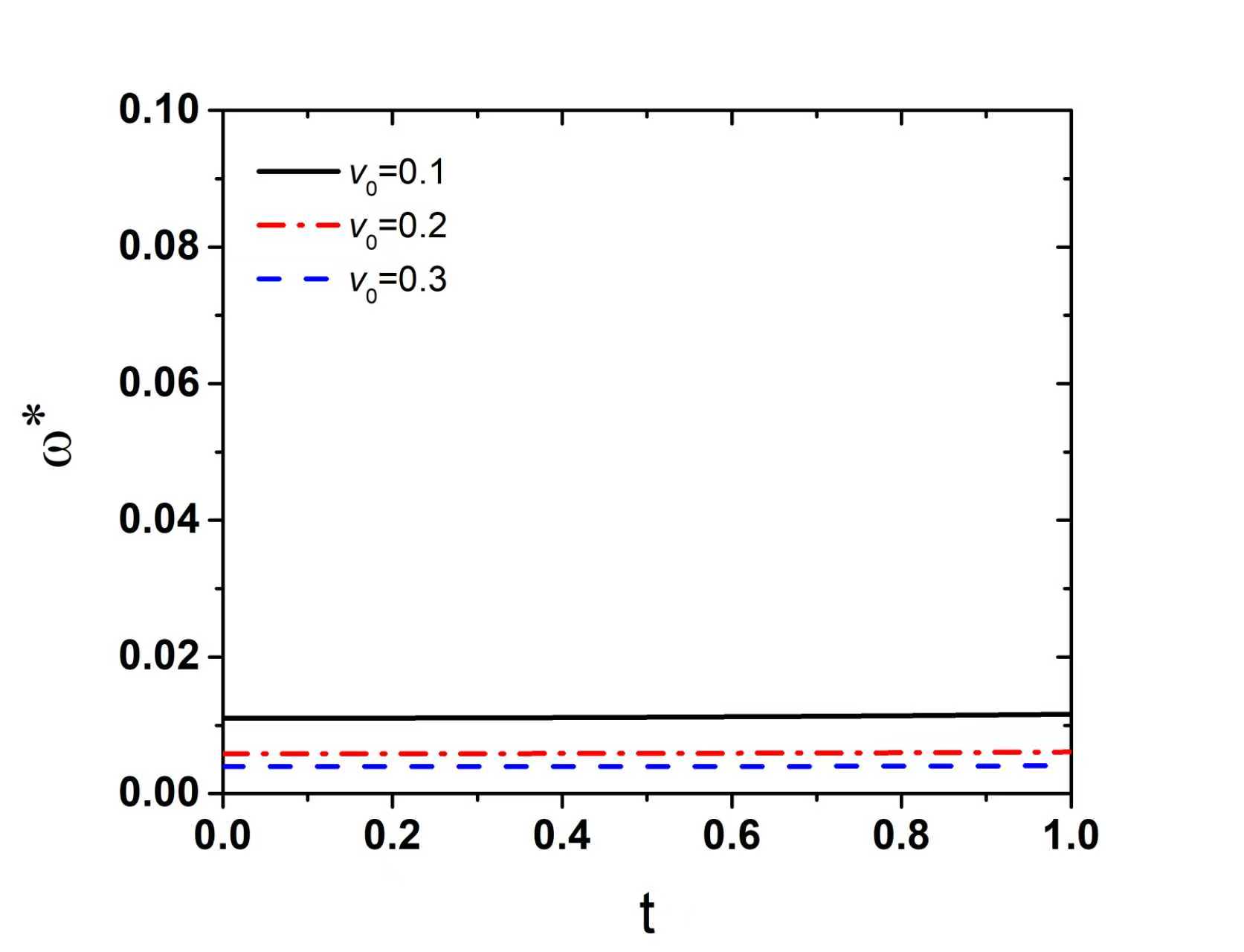}}
  \caption{Different $v_0$.}\label{fig:v0}
\end{figure}

Finally, we analyze the impact of the initial volatility level, $v_0$, on the optimal investment strategy, as illustrated in Figure \ref{fig:v0}. Given the multi‐dimensional nature of the volatility structure, we focus specifically on $v_0$ to isolate its effect, holding all other parameters constant. Numerical results are compared across $v_0=0.1$, $0.2$, and $0.3$. A pronounced inverse relationship is observed: as $v_0$ rises, the optimal stock proportion declines significantly. This pattern is consistent with portfolio theory: higher initial volatility elevates the perceived risk of equity holdings, prompting risk-averse investors to reduce their exposure to stocks. The strong sensitivity to $v_0$ underscores the critical role of volatility expectations in shaping portfolio choice. It also highlights that, even in a market with liquidity frictions and transaction costs, traditional risk-return considerations remain a dominant driver of investor behavior.

\section{Conclusion}
This study has developed and analyzed a portfolio selection model within a Merton continuous-time framework, incorporating several key features of real-world financial markets. The model accounts for both exogenous proportional transaction costs and endogenous liquidity risk, the latter modeled as a stochastic process, and examines their interplay in an incomplete market setting. A two-factor stochastic volatility structure is integrated, where volatility follows a mean-reverting process with a stochastic long-term mean, allowing for a more realistic representation of market dynamics.

In selecting the investor’s utility function, an option-implied approach was employed to align with observed market behavior. The S-shaped utility function, grounded in prospect theory, was identified as the most statistically suitable specification. To address its inherent non-concavity, a concave envelope transformation was applied, ensuring the resulting optimization problem remains well-posed and computationally tractable. The resulting high-dimensional nonlinear HJB equation was solved using a deep learning-based policy iteration scheme. By leveraging PINNs, the algorithm effectively learned both the value and policy functions, overcoming challenges associated with dimensionality and nonlinearity.

A series of numerical experiments were conducted, validating the robustness and efficiency of the proposed method. More importantly, these experiments quantitatively assessed the distinct influences of exogenous and endogenous transaction costs, as well as stochastic volatility factors, on optimal portfolio allocation. The results underscore the importance of incorporating market frictions and behavioral preferences into dynamic asset allocation models, offering both theoretical insight and practical relevance for portfolio management under realistic financial conditions.

\begin{appendices}
\section{Utility functions in Section 3.4}\label{Utilityfct}
Below are several classical utility functions frequently referenced in economic analysis. We present their functional forms along with  relative risk aversion (RRA):

Example 1:  Power utility

The Power utility can be defined as:
\begin{equation}
U(W) = \frac{1}{1 - k}W^{1 - k},
\end{equation}

Additionally, the relative risk aversion (RRA) of the Power utility can be calculated based on the following formula:
\begin{equation}
RRA(W) = -\frac{U''(W)}{U'(W)}W = k.
\end{equation}

Example 2:  Exponential utility

The Exponential utility can be defined as:
\begin{equation}
U(W) = -\frac{e^{-kW}}{k},
\end{equation}

Additionally, the relative risk aversion (RRA) of the Exponential utility can be calculated based on the following formula:
\begin{equation}
RRA(W) = -\frac{U''(W)}{U'(W)}W = kW.
\end{equation}

Example 3: HARA utility

The HARA utility can be defined as:
\begin{equation}
U(W) = \frac{1}{k_1 - 1} \left[ k_1 W + k_2 \right]^{1 - \frac{1}{k_1}},
\end{equation}

Additionally, the relative risk aversion (RRA) of the HARA utility can be calculated based on the following formula:
\begin{equation}
RRA(W) = -\frac{U''(W)}{U'(W)}W = \frac{W}{k_1 W + k_2}.
\end{equation}

Example 4: Log plus power utility

The Log plus power utility is a combination of the logarithmic utility and the power utility, which can be defined as:
\begin{equation}
U(W) = k_1 \log W + \frac{1}{k_2} W^{k_2},
\end{equation}

Additionally, the relative risk aversion (RRA) of the Log plus power utility can be calculated based on the following formula:
\begin{equation}
RRA(W) = -\frac{U''(W)}{U'(W)}W = \frac{k_1 + (1 - k_2) W^{k_2}}{k_1 + W^{k_2}}.
\end{equation}

Example 5: Linear plus exponential utility

The Linear plus exponential utility is a combination of the linear utility and the exponential utility, which can be defined as:
\begin{equation}
U(W) = k_1 W - \frac{1}{k_2} e^{-k_2 W},
\end{equation}

Additionally, the relative risk aversion (RRA) of the Linear plus exponential utility can be calculated based on the following formula:
\begin{equation}
RRA(W) = -\frac{U''(W)}{U'(W)}W = \frac{k_2 W e^{-k_2 W}}{k_1 + e^{-k_2 W}}.
\end{equation}

\section{Components of neural networks}\label{Neutral work}
For each layer, the components for the values network $Q_{\phi}=f_{3}^{Q} \circ\tanh \circ f_{2}^{Q}\circ\tanh \circ f_{1}^{Q}$ are defined as

$$ f_{3}^{Q}:\mathbb{R}^{N^{Q}}\rightarrow\mathbb{R},\quad x\mapsto W_{3}^{Q}x+b_{3}^{Q}, $$ 

$$ \tanh:\mathbb{R}^{N^{Q}}\rightarrow\mathbb{R}^{N^{Q}},\quad x\mapsto\tanh(x), $$ 

$$ f_{2}^{Q}:\mathbb{R}^{N^{Q}}\rightarrow\mathbb{R}^{N^{Q}},\quad x\mapsto W_{2}^{Q}x+b_{2}^{Q}, $$ 

$$ \tanh:\mathbb{R}^{N^{Q}}\rightarrow\mathbb{R}^{N^{Q}},\quad x\mapsto\tanh(x), $$ 

$$ f_{1}^{Q}:\mathbb{R}^{5}\rightarrow\mathbb{R}^{N^{Q}},\quad x\mapsto W_{1}^{Q}x+b_{1}^{Q}, $$ 

 and the components for the control network $\omega_{\psi}=\operatorname{sigmoid}\circ f_{3}^{\omega}\circ\tanh\circ f_{2}^{\omega}\circ\tanh\circ f_{1}^{\omega}$ are given by

$$ \operatorname{sigmoid}:\mathbb{R}\rightarrow\mathbb{R},\quad x\mapsto\frac{1}{1+\mathrm{e}^{-x}}, $$ 

$$ f_{3}^{\omega}:\mathbb{R}^{N^{\omega}}\rightarrow\mathbb{R},\quad x\mapsto W_{3}^{\omega}x+b_{3}^{\omega},$$ 

$$ \tanh:\mathbb{R}^{N^{\omega}}\rightarrow\mathbb{R}^{N^{\omega}},\quad x\mapsto\tanh(x), $$ 

$$ f_{2}^{\omega}:\mathbb{R}^{N^{\omega}}\rightarrow\mathbb{R}^{N^{\omega}},\quad x\mapsto W_{2}^{\omega}x+b_{2}^{\omega},$$ 

$$ \tanh:\mathbb{R}^{N^{\omega}}\rightarrow\mathbb{R}^{N^{\omega}},\quad x\mapsto\tanh(x), $$ 

$$ f_{1}^{\omega}:\mathbb{R}^{5}\rightarrow\mathbb{R}^{N^{\omega}},\quad x\mapsto W_{1}^{\omega}x+b_{1}^{\omega}. $$ 

Here, $N_{Q}$ and $N_{\omega}$ denote the hidden layer sizes for each network. Unless specified otherwise, these values are set to 64. Note that both networks are optimized using the LBFGS optimizer, with a learning rate set to $10^{-1}$.

\end{appendices}

\bibliographystyle{apalike}
 \bibliography{ref}

\end{document}